\documentclass[3p,12pt]{elsarticle}
\biboptions{sort&compress}
\usepackage{amssymb}
\usepackage{color,slashed,mathrsfs}

\newcommand{\be}{\begin{equation}}
\newcommand{\ee}{\end{equation}}
\newcommand{\bea}{\begin{eqnarray}}
\newcommand{\eea}{\end{eqnarray}}
\newcommand{\vecq}{{\vec q}}
\newcommand{\vecp}{{\vec p}}

\newcommand{\ep}{{\varepsilon}}
\newcommand{\Imx}{{\rm Im}}

\definecolor{red}{rgb}{0.8,0,0}
\definecolor{violet}{rgb}{0.4,0,0.4}
\definecolor{green}{rgb}{0,0.5,0.0}
\definecolor{navy}{rgb}{0.0,0.0,0.6}
\definecolor{orange}{rgb}{0.8,0.2,0.0}
\definecolor{blue}{rgb}{0.3,0.0,0.8}

\newcommand{\ie}{\emph{i.e.}}

\begin{document}
\journal{Nucl. Phys. A, vol. 897,  Jan. 2013, pp. 62-69}

\begin{frontmatter}
\title{ {\bf  Axions  from cooling compact stars: pair-breaking processes}}
\author{\normalsize{  Jochen Keller 
                         and Armen  Sedrakian
}
}
\address{
 Institute for Theoretical Physics, J. W.\ Goethe-University, D-60438 Frankfurt am Main, Germany
}

\date{31 May 2012}

\begin{abstract}
  Once formed in a supernova explosion, a neutron star cools rapidly
  via neutrino emission during the first $10^4$-$10^5$ yr of its
  life-time. Here we compute the axion emission rate from baryonic
  components of a star at temperatures below their respective critical
  temperatures $T_c$ for normal-superfluid phase transition. The axion
  production is driven by a charge neutral weak process, associated
  with Cooper pair breaking and recombination. The requirement that
  the axion cooling does not overshadow the neutrino cooling puts a
  lower bound on the axion decay constant $f_a > 6\times 10^{9}\, T_{c\, 9}^{-1}$
  GeV, with $T_{c\,9} = T_c/10^{9}$ K. This translates into a upper
  bound on the axion mass $m_a < 10^{-3}\, T_{c\, 9}$ eV.
\end{abstract}

\begin{keyword}
 Axions \sep Neutron stars 
\MSC[2011]                   14.80.Va   \sep  97.60.Jd     
\end{keyword}
\end{frontmatter}

\section {Introduction}\label{introduction}

CP-violation in the strong sector of the Standard Model arises due to a
topological interaction term in the QCD Lagrangian
\be
\label{thetaaction}
\mathscr{L}_\theta={g^2\theta\over 32\pi^2}
 \,F^a_{\mu\nu}\tilde F^{\mu\nu a}, 
\ee
where $F_{\mu\nu}^a
= \partial_{\mu}A_{\nu}-\partial_{\nu}A_{\mu}+gf^{abc}A_{\mu b} A_{\nu
  c}$ is the gluon field strength tensor, $g$ is the strong coupling
constant, $\tilde F^a_{\mu\nu} =
\epsilon_{\mu\nu\lambda\rho} F^{\lambda\rho a}/2$, $f^{abc}$ are the
structure constants of $SU(3)$ group, $\theta$ is the parameter which
parametrizes the non-perturbative vacuum states of QCD
$\vert\theta\rangle = \sum_n \exp(-in\theta)\vert n\rangle$, where $n$
is the winding number characterizing each distinct state, which
is not connected to another by any gauge
transformation~\cite{'tHooft:1976up}.  The QCD action changes by
$2\pi$ under the shift $\theta\to\theta+2\pi$, \ie, $\theta$ is a
periodic function with a period of $2\pi.$ In presence of quarks the
physical parameter is not $\theta$, but
\be
\label{thetabar}
\bar\theta = \theta + \arg\det m_q , \ee where $m_q$ is the matrix of
quark masses. Experimentally, the upper bound on the value of this
parameter is $\bar\theta\lesssim 10^{-10}$, which is based on the
measurements of the electric dipole moment of neutron $d_n<6.3\cdot
10^{-26}e$ cm~\cite{Baker:2006ts}.  The smallness of $\bar\theta$ is
the strong CP problem:  the Standard Model does not provide any
explanation on why this number should not be of order unity.

An elegant solution to the strong CP-problem is provided by the
Peccei-Quinn
mechanism~\cite{Peccei:1977hh,Weinberg:1977ma,Wilczek:1977pj}.  This
solution amounts to introducing a global $U(1)_{PQ}$ symmetry, which
adds an additional anomaly term to the QCD action proportional to the
axion field $a$. This term acts as a potential for the axion field and
gives rise to an expectation value of the axion field $\langle a
\rangle \sim -\bar\theta$.  The physical axion field is then $a-
\langle a \rangle $, so that the undesirable $\theta$ term in the
action is replaced by the physical axion field. The axion is the
Nambu-Goldstone boson of the Peccei-Quinn $U(1)_{PQ}$ symmetry
breaking~\cite{Weinberg:1977ma,Wilczek:1977pj},
 and its effective Lagrangian has the form \be \mathscr{L}_a =
-\frac{1}{2}\partial_{\mu} a\partial^{\mu} a +
\mathscr{L}_{int}(\partial_{\mu} a,\psi), \ee where the second term
describes the coupling of the axion to fermion fields ($\psi$) of the
Standard Model.

There are ongoing experimental searches for the axion and the
cosmology and astrophysics provide strong complementary constraints.
Because axions can be effectively produced in the interiors of stars
they act as an additional sink of energy. The requirements that the
energy loss from a star is consistent with the astrophysical
observations place lower bounds on the coupling of axions to the
Standard Model particles, and hence on the Peccei-Quinn symmetry
breaking scale~\cite{Raffelt:2006cw,Raffelt:2011ft}. The latter limit
translates into an upper limit on the axion mass.  Such arguments have
been applied to the physics of supernova explosions
~\cite{Brinkmann:1988vi,Burrows:1988ah,Raffelt:1993ix,Janka:1995ir,Hanhart:2000ae}
and white dwarfs~\cite{Corsico:2012ki}. In the case of supernova
explosions the dominant energy loss process is the emission of an
axion in the nucleon ($n$) bremsstrahlung $n+n\to n+n+a$. The same
process was considered earlier by Iwamoto as a cooling mechanism for
mature neutron stars, \ie, neutron stars with core temperature in the
range $10^8-10^9$~K~\cite{Iwamoto:1984ir}.  The implications of the
axion emission by the modified nucleon bremsstrahlung, as
calculated in Ref.~\cite{Iwamoto:1984ir}, on the cooling of neutron
stars were briefly discussed in Ref.~\cite{Umeda:1997da}. However, it
is now well established through cooling simulations of compact
stars~\cite{Page:2005fq,Yakovlev:2007vs,Sedrakian:2006mq}, that their
neutrino cooling era, which spans the time period $t\le 10^4-10^5$ yr
after birth, is strongly affected by the neutrino emission from its
superfluid phases due to the process of neutrino emission by
Cooper-pair
breaking~\cite{Flowers:1976ux,Yakovlev:1998wr,Kaminker:1999ez,Sedrakian:2006ys,Kolomeitsev:2008mc,Sedrakian:2012ha}.

In this article we compute the rate at which the superfluid phases of
a neutron star lose their energy by axion emission via the processes
of Cooper pair breaking and recombination.  This work concentrates on
the baryonic interiors of compact stars and considers for the sake of
simplicity $S$-wave superfluids.  The inner crusts and the baryonic
core of a neutron star features iso-triplet spin-0 $S$-wave
superfluids; in addition the core may contain spin-1, $P$-wave neutron
superfluid~\cite{Pwave1,Pwave2,Pwave3} and at high densities,
iso-singlet, spin-1, neutron-proton $D$-wave
superfluid~\cite{Alm:1996zz}.  The extension of the present work to
$P$-wave and $D$-wave superfluids is straightforward. We use natural
units, $\hbar = c = k_B=1 $.

\section{Currents, matrix elements and emissivity}

The coupling of axion to baryonic fields is described by the following
interaction Lagrangian
\be 
\mathscr{L }_{int} = \frac{1}{f_a} B^{\mu}  L_{\mu},
\ee
where $f_a$ is the axion decay constant, the baryon and axion currents 
are given by
\be
B^{\mu} = C_a \bar\psi\gamma^{\mu}\gamma_5\psi,\quad \quad 
L_{\mu} = \partial_{\mu}\phi,
\ee
where $C_a$ are model dependent coupling constants of order of
unity. In the case of a multicomponent baryonic system the baryon
current contains a sum over all components.  The squared matrix element 
for the process of axion emission is then given by
\bea
\vert{\cal M}_{a}\vert^2 =\frac{1}{2}f_a^{-2}
  (B^{\mu} B^{\nu\dagger}) (L_{\mu}L_{\nu}^{\dagger}) .
\eea
The energy radiated per unit time in axions (axion emissivity) is
given by the phase-space integral over the probability of the process
of emission
\be\label{emiss1}
\epsilon_a = -f_a^{-2}
\int\frac{d^3q}{(2\pi)^32\omega} ~\omega g(\omega) q_{\mu}q_{\nu}  \Imx \Pi_a^{\mu\nu}(q),
\ee
where $q$ and $\omega$ are the axion momentum and energy. Here 
we defined the polarization tensor of baryonic matter
\be
\Imx \Pi^{\mu\nu}(\omega,\vecq) = 
 \frac{1}{2}\sum_{n}(B_{\mu}B_{\nu}^{\dagger}) \delta^4(q-\sum_i p_i),
\ee
where the $i$ sum is over the four-momenta of the baryons. Upon carrying
out the angular integral in Eq. (\ref{emiss1}) we write the
emissivity in terms of a one-dimensional integral
\be\label{axion_emiss}
\epsilon_a = -\frac{f_a^{-2}}{4\pi^2} \int_0^{\infty}
d\vert\vecq\vert ~ \vecq^2 g(\omega) \kappa_a(q) ,
\ee
where the contraction of the axion currents with the baryonic
polarization tensor is given by 
\be\label{eq:kappa} 
\kappa_a(q) = q_{\mu}q_{\nu} \Imx \Pi_a^{\mu\nu}(q).
\ee
So far the expression for the axion emissivity is completely
general; we will need to compute the polarization tensor of
baryonic matter for the process of interest.

\section{Polarization tensor of superfluid baryon matter}

At sufficiently low densities and temperatures baryonic matter forms a $^1S_0$ pair
condensate. In compact stars this is the case for all baryons except
neutrons, which may form $P$-wave superfluid at densities 
at and above the saturation density. To describe the response of baryonic
matter to the axion field we use the methods developed for the
description of neutrino interactions in 
Refs.~\cite{Sedrakian:2006ys,Sedrakian:2012ha} (see also 
~\cite{Kolomeitsev:2008mc,Baldo:2011nc}).
The  imaginary-time momentum-space  correlators are given
by the $2 \times 2$ Nambu-Gor'kov matrix
\bea \label{eq:propmatrix} {\cal G}_{\sigma,\sigma'}(i\omega_n,\vecp)
&=& \left(\begin{array}{cc} \hat G_{\sigma\sigma'}(i\omega_n,\vecp)
    &\hat
    F_{\sigma\sigma'}(i\omega_n,\vecp)   \\
    \hat F^+_{\sigma\sigma'}(i\omega_n,\vecp) & \hat
    G^+_{\sigma\sigma'}(i\omega_n,\vecp)
\end{array}
\right).
\eea
The elements of the matrix are time-ordered  correlators of the baryon
field $\psi_B$ and $\psi^{\dagger}_B$; in the frequency-momentum
domain these are given by 
\bea \label{Prop_P}
\hat G_{\sigma\sigma'}(i\omega_n,\vecp) 
&=& \delta_{\sigma\sigma'}\left(
\frac{u_p^2}{i\omega_n-\ep_p} +\frac{v_p^2}{i\omega_n+\ep_p} \right),\\
    \label{Prop_F}
    \hat F_{\sigma\sigma'}(i\omega_n,\vecp) &=& - i\sigma_y
    u_pv_p\left(\frac{1}{i\omega_n-\ep_p}-\frac{1}{i\omega_n+\ep_p}\right),
    \eea where $ F_{\sigma\sigma'}^{+}(i\omega_n,\vecp) =
    F_{\sigma\sigma'}(i\omega_n,\vecp)$, $\omega_n = (2n+1)\pi T$ is
    the fermionic Matsubara frequency, $\sigma_y$ is the $y$ component
    of the Pauli-vector in spin space, $ u_p^2
    =(1/2)\left(1+{\xi_p}/{\ep_p}\right) $ and $u_p^2 +v_p^2 = 1$
    $\ep_p = \sqrt{\xi_p^2+\Delta_p^2}$ is the single particle energy in the
    paired state, with $\Delta_p$ being the (generally momentum- and
    energy-dependent) gap in the quasiparticle spectrum and $\xi_p =
    v_F (p-p_F)$ is the single-particle spectrum in the normal state,
    where $v_F$ and $p_F$ are the (effective) Fermi velocity and
    momentum. 
   The propagator for the holes is defined as $\hat G_{\sigma\sigma'}^{+}
    (i\omega_n,\vecp) = \hat G_{\sigma\sigma'}(-i\omega_n,-\vecp) $.
    For an $S$-wave condensate we have  $\hat
    G_{\sigma\sigma'}(i\omega_n,\vecp)=\delta_{\sigma\sigma'}
    G(i\omega_n,\vecp)$, $\hat F_{\sigma\sigma'}(i\omega_n,\vecp) = -
    i\sigma_yF(i\omega_n,\vecp)$, {\it etc}.
\begin{figure*}[t]
\begin{center}
\includegraphics[height=1.2cm,width=12cm]{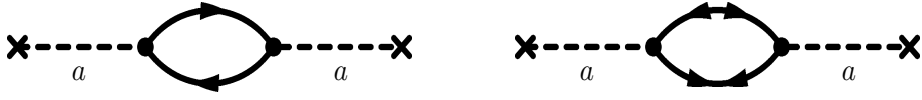}
\end{center}
\caption[] { The two diagrams contributing to the polarization tensor of
 baryonic matter,  which defines  the axion emissivity.
The ``normal'' baryon propagators for particles (holes) are shown
by single-arrowed lines directed from left to right (right to left).
 The double arrowed lines correspond to the ``anomalous''
propagators $F$ (two incoming arrows) and $F^+$ (two outgoing
arrows).  The horizontal dashed lines represent the axion $a$.  }
\label{fig1}
\end{figure*}
The polarization tensor of a superfluid obtains contributions from
four distinct diagrams that can be formed from the normal and
anomalous propagators with four distinct effective vertices. However,
for the axial vector perturbations the vertices are not renormalized
in the medium and, therefore, one proceeds with the bare vertices, in
which case the number of the distinct contributions to the
polarization tensor reduces to a sum of two admissible bare loops (see
Fig.~\ref{fig1}) \be \label{A} {\cal A } ^{\cal T}(q)\equiv
\Pi_{GG}(q) -{\cal T} \Pi_{FF}(q), \ee where ${\cal T} = \pm1 $ is the
time reversal operator and \bea \label{eq:pi1} \Pi_{GG}(q)=
T\int\frac{d\vecp}{(2\pi)^3}\sum_{ip_n} G(p)G(p+q) =
\frac{\nu(0)}{4}\int_{-1}^{1}\! dx (G * G), 
\eea 
where $T$ is the temperature, $p\equiv (ip_n,\vecp)$ with $p_n$ being the fermionic
Matsubara frequency, $\nu(0) = m^*p_F/\pi^2$ is the density of states
of neutrons with $m^*$ being their effective mass, which may include
the wave function renormalization (the so-called $E$-mass), $x$ is the
cosine of the angle formed by the vectors $\vecq$ and $\vecp$. The
convolution is defined as 
\be (G * G)=T\int_{-\infty}^{\infty}d\xi_p
\sum_{ip_n} G(p)G(p+q), 
\ee with a similar expression for $(F *
F)$. Carrying out the Matsubara sums appearing in the convolutions
we find
\bea {\cal A}^{\pm} (q) &=& -2\nu(0)\overline{
  \left[\frac{x\delta}{1-x\delta}+\frac{(1\pm1)}{2}\right]{(F*F^+)}},
\eea 
where $\overline{(\dots)} \equiv (1/2)\int_{-1}^1 \, dx(\dots) $,
$x \equiv \hat q\cdot \hat p$ and $\delta = \vert q \vert v_F/\omega$
is a small parameter of the theory.  To leading order in $\delta$
parameter the imaginary part of the $(F*F)$ convolution can be
computed analytically 
\bea\label{ImFF} \Imx (F*F^+)_0& = & \frac{\pi\Delta^2}{\omega}
\frac{\theta(\omega-2\Delta)}{\sqrt{\omega^2-4\Delta^2}} \tanh
\left(\frac{\omega}{4T}\right),
\eea whereas the real part follows from the dispersion relation 
\bea
{\rm Re} (F*F^+)_0&=& \frac{1}{\pi}\int_{-\infty}^{\infty}
\frac{d\omega'}{\omega-\omega'}\Imx (F*F^+)_0(\omega').  \eea For the
contraction (\ref{eq:kappa}) we obtain \bea \kappa_a(q) &=& \omega^2 \Imx
\Pi^{00}(q)-q_{i}\omega \Imx \Pi^{i0}(q) -\omega q_{j} \Imx
\Pi^{0j}(q) +\frac{\vecq^2}{3} \Imx \Pi^{ii}(q), \eea where, to the
accuracy we are working, $\Imx \Pi_A^{00}(q) = {\cal A}^+ v_F^2$,
$\Imx \Pi^{ii}(q) = 3{\cal A}^-(q)$, $q_j\Imx \Pi^{0j}(q) =
\tilde{\mathcal{A}}^-\,qv_F$ and $q_i\Imx \Pi^{i0}(q) =
\tilde{\mathcal{A}}^+\,qv_F. $ Here $\tilde{\cal A}^{\pm} \equiv {\cal
  A}^\pm(\hat q\cdot \hat p)$ are the first moments of ${\cal A}^{\pm}$
integrals with respect to the cosine of the angle formed by the
vectors $\vecq$ and $\vecp$.  The leading order in $\delta$ expansion of the
contraction gives \bea \label{kappa_final} \kappa_a(q) &=& -2\nu(0)
\left[\omega^2 v_F ^2 \left(1+\frac{\delta^2}{3}\right) - 
  \frac{2\omega^2\delta^2}{3} 
  + \frac{\vecq^2 \delta^2}{3}\right]  {\rm Im }(F*F^+)_0\nonumber\\
&=&-\frac{4\nu(0)}{3}\vecq^2v_F^2 {\rm Im }(F*F^+)_0 + O(\delta^4).
\eea 
Eq. (\ref{kappa_final}) completes our evaluation of the baryonic
polarization tensor and its contraction with axionic current, which is
the key input for the computation of the axion emissivity.  Note that to
$O(\delta^4)$ accuracy the $(F*F^+)$ convolution appears at its lowest 
($\delta = 0$) order. More accurate
evaluation is not required, because of the major uncertainty in the
coupling strength of axions to the Standard Model fermions as well as
other large uncertainties in the physics of cooling neutron stars.

\section{Axion emissivity}
The axion emissivity is obtained on substituting Eqs.  (\ref{ImFF})
and (\ref{kappa_final}) in Eq. (\ref{axion_emiss})
\bea\label{eq:axion1} \epsilon_a &=& \frac{ 8 }{3\pi} \, f_a^{-2}
\,\nu(0)\, v_F^2 \, T^5 \, I_a, \eea where \bea I_a =
z^5\int_1^{\infty}\!\! dy ~ \frac{y^3}{\sqrt{y^2-1}} f_F\left(z
  y\right)^2, \eea $z= \Delta(T)/T$ and $f_F (x) = [1+\exp(x)]^{-1}$
is the Fermi distribution function.  The $T^5$ scaling of the
emissivity is understood as follows. The integration over the phase
space of neutrons carries a power of $T$, since for degenerate
neutrons the phase-space integrals are confined to a narrow strip
around the Fermi surface of thickness $T$. The axion is emitted
thermally and being relativistic contributes a factor $T^3$ to the
emissivity. The one power of $T$ from the energy of the axion and the
inverse one power of $T$ from the energy conserving delta function
cancel. The transition matrix element is proportional to the
combinations of $u_p$ and $v_p$ amplitudes, which are dimensionless,
but contain {\it implicit} temperature dependence due to the
temperature dependence of the gap function. This dependence is not
manifest in Eq.~(\ref{eq:axion1}), \ie, was absorbed in the definition
of the integral $I_a$. Thus, the explicit temperature dependence of
the axion emission rate Eq.~(\ref{eq:axion1}) is $T^5$.  In the cgs
units the axion emissivity Eq.~(\ref{eq:axion1}) is \bea \epsilon_a&=&
1.06 \times 10^{21} \left(\frac{10^{10} \textrm{GeV}}{f_a}\right)^2
\left(\frac{m^*}{m}\right)^2 \left(\frac{v_F}{c}\right)^3
\left(\frac{T}{10^9\textrm{K}}\right)^5 \, I_a\, \textrm{erg} ~
\textrm{cm}^{-3} ~ \textrm{s}^{-1}, \eea where two powers of $v_F/c$
arise from the small momentum transfer expansion and one power - from
the density of states.  At temperatures of order the critical
temperature $T_c\simeq 10^9$~K the superfluid cools primarily by
emission of neutrinos via the pair-breaking processes driven by the
axial-vector currents (we continue to assume that potential fast
cooling via direct Urca processes is prohibited).  The emissivity of
this processes in the case of $^1S_0$-wave superfluid is given
by~\cite{Flowers:1976ux,Yakovlev:1998wr,Kolomeitsev:2008mc} \be
\epsilon_{\nu}=\frac{4G_F^2g_A^2}{15\pi^3} \zeta_A\nu(0) {v_F}^2T^7
I_{\nu}, \ee where $G_F$ is the weak Fermi coupling constant, $g_A =
1.25$ is the axial-vector current coupling constant, $\zeta_A = 6/7$
and 
\be I_{\nu} = z^7 \int_1^{\infty} dy \frac{y^5}{\sqrt{y^2-1}} f_F
\left(z y\right)^2 .  
\ee 
We now require that the axion luminosity
does not exceed the neutrino luminosity, \ie, 
\bea\label{eq:ratio}
\frac{\epsilon_a}{\epsilon_{\nu}} &=& 
\frac{10\pi^2}{f_a^2 G_F^2g_A^2\zeta_A }  \frac{I_a}{I_\nu}< 1.
\eea 
Substituting the the free-space value of the axial vector coupling $g_A=
1.25$  and introducing $r(z) \equiv z^2(I_a/I_{\nu}) $
we transform Eq. (\ref{eq:ratio})
\bea \label{eq:ratio2}
\frac{\epsilon_a}{\epsilon_{\nu}} &=&
\frac{59.2}{ f_a^{2} G_F^2 \Delta(T)^2}~r(z).
\eea
Not far from the critical temperature 
$\Delta (T) \simeq 3.06T_c\sqrt{1-T/T_c}$, 
which translates into $z = 3.06\, t^{-1}\sqrt{1-t}$,  
where $t = T/T_c$. Numerical evaluations of the integrals provides the
following values  $r(0.5) = 0.07$,
$r(1) = 0.26$,  $r(2) = 0.6$ and 
asymptotically  $r(z) \to 1$ for $z\gg 1$.
Substituting the value of the Fermi coupling constant $ G_F =
1.166\times 10^{-5}$ GeV$^{-2}$ in Eq.~(\ref{eq:ratio2}) and 
noting that $r(z)\le 1$ , we finally obtain 
\be\label{eq:fbound} f_a > 5.92 \times 10^{9}\, \textrm{GeV} \,
\left[\frac {0.1~\textrm{MeV}}{\Delta(T)}\right]
\ee
which translates into an upper bound on the axion mass 
\be \label{eq:mbound} m_a = 0.62 \times
10^{-3}\, \textrm{eV}\, \left(\frac{10^{10}
    \textrm{GeV}}{f_a}\right)\le 1.05 \times 10^{-3}\, \textrm{eV}\,
\left[\frac {\Delta(T)}{0.1~\textrm{MeV}}\right].  
\ee 
The bound (\ref{eq:fbound}) can be written in terms of the critical
temperature by noting that $\Delta (T) \simeq T_c$ in the temperature range $0.5\le
t< 1$ of most interest.

\section{Discussion}

The neutrino cooling era of compact stars, which spans the time-period
$ t\le 10^4-10^5$ yr after their birth in supernova explosions is a
sensitive probe of the particle physics of their interiors. If one
assumes that there are no rapid channels of cooling in neutron stars,
\ie, deconfined quarks, above Urca threshold fractions of protons or
hyperons (all of which lead to a rapid Urca cooling), then neutron
stars cool primarily by neutrino emission in Cooper pair-breaking
processes in baryonic superfluids. Here we have shown that if axions
exist in Nature, the neutron stars must cool via axion emission in
Cooper pair-breaking processes, whose axion emission rate scales as
$T^5$. This scaling differs from the $T^7$ scaling of the counterpart
neutrino processes. The difference arises from the different phase
spaces required for an axion and a pair of neutrinos and is
independent of the baryonic polarization tensor. Note also that the
rate of axion emission from a $P$-wave superfluid will differ from the
$S$-wave rate, derived above, by a factor $O(1)$ and, therefore, will
not change quantitatively the obtained bounds on the axion parameters.

If the Standard Model physics provides a consistent explanation of the
data on the cooling of neutron stars, one can place a lower bound on
the axion decay constant (the breaking scale of the Peccei-Quinn
symmetry). Our calculations show that this bound is given by $f_a >
5.92\times 10^{9} T_{c\,9}^{-1}$ GeV, where $T_{c\,9} $ is the
magnitude of the critical temperature in units of $10^9$ K.  This
translates into an upper bound on the axion mass $m_a \le 10^{-3}\,
T_{c\,9}$ eV. Similar bounds were obtained previously by
Iwamoto~\cite{Iwamoto:1984ir} ($f /10^{10}\textrm{GeV} > 0.3$) from
comparison of the rates of axion bremsstrahlung and modified Urca
neutrino emission by mature neutron stars, and by Umeda et
al~\cite{Umeda:1997da} ($f /10^{10}\textrm{GeV} > 0.1-0.2$) from fits
of cooling simulations to the PSR 0656+14 data.\footnote{Note that
  Eq. (9) of this work is incorrect, which could be the reason why the
  limits on the axion's mass reported in their Figs. 2-4 are by an
  order of magnitude larger compared to those derived in this work.}
Our lower bound on $f_a$ is somewhat larger than the one that follows
form the requirement that the axions do not ``drain'' too much energy
from supernova process so that it
fails~\cite{Raffelt:2006cw,Brinkmann:1988vi,Burrows:1988ah,Raffelt:1993ix,Janka:1995ir,Hanhart:2000ae}. Furthermore,
our bound sensitively depends on the pairing gap in baryonic
superfluids, whose magnitude and density dependence are not
well-known.
 The physical implications of the bound (\ref{eq:fbound})
  can be fully explored with numerical simulations of neutron star
  cooling. Targeted fits to a specific object exhibiting slow cooling
  would be more useful than fits to the entire population of neutron
  stars with measured $X$-ray fluxes. Examples, of such fits were
  carried out, e. g., in the case of the neutron star in CAS
  A~\cite{Page:2010aw,Blaschke:2011gc}.  The accuracy of the
  predictions will be limited by the uncertainties in the physical
  input in cooling simulations and uncertainties inherent to the
  interpretation of the data.  For stars in the neutrino cooling era
  the dissipative heating processes are unimportant. The potential
  sources of uncertainty are well documented in the
  literature~\cite{Page:2005fq,Yakovlev:2007vs,Sedrakian:2006mq} and
  include (a) the rate of neutrino/axion emission, in particular, its
  dependence on the magnitude of the pairing gap; (b) the composition
  of the surface layers; (c) the influence of the star's $B$-field on
  the photon and neutrino emission processes; (d) the estimate of the
  age of any given object. We anticipate that the error bars on the
  bounds should remain within a factor of few and the accuracy of the
  bounds can be improved with the help of numerical simulations of
  cooling compact stars.

\end{document}